\documentclass[showpacs,preprintnumbers,amsmath,amssymb,aps]{revtex4}

\usepackage{graphicx}% Include figure files
\usepackage{dcolumn}% Align table columns on decimal point
\usepackage{bm}% bold math

\begin{document}

\title{Minority Game With Peer Pressure}

\author{H. F. Chau, F. K. Chow and K. H. Ho}
\affiliation{
  Department of Physics, University of Hong Kong, Pokfulam Road, Hong Kong
}
\date{\today}

\begin{abstract}
To study the interplay between global market choice and local peer pressure,
we construct a minority-game-like econophysical model. In this so-called
networked minority game model, every selfish player uses both the historical
minority choice of the population and the historical choice of one's
neighbors in an unbiased manner to make decision. Results of numerical
simulation show that the level of cooperation in the networked minority
game differs remarkably from the original minority game as well as the
prediction of the crowd-anticrowd theory. We argue that the deviation from
the crowd-anticrowd theory is due to the negligence of the effect of a four
point correlation function in the effective Hamiltonian of the system. 
\end{abstract}


\maketitle

\section{Introduction}

Minority Game (MG) \cite{MG1,MG2} --- a game proposed by Challet and Zhang
under the inspiration of the El Farol bar problem \cite{Elf} --- is a simple
model showing how selfish players cooperate with each other in the absence
of direct communication. It succinctly captures the self-organizing global
cooperative behavior which is ubiquitously found in many social and
economic systems.

In MG, $N$ inductive reasoning players have to choose one out of two choices
independently in each turn. Based only on certain commonly available global
information, each player has to decide one's choice by means of his/her
current best working strategy or mental model. Those who end up in the
minority side (i.~e.,~the choice with the least number of players) win.
Although its rules are remarkably simple, MG exhibits very rich
self-organized collective behavior \cite{MGNu1,MGNu2,MGNu3}. Moreover, the
dynamics of MG can be explained by the so-called crowd-anticrowd theory
\cite{CAC1,CAC2,CAC3} which stated that fluctuations arisen in the MG is
resulted from the interaction between crowds of like-minded agents and their
anti-correlated partners.
 
However, in the real world, people usually do not only consider the global
information when they make decisions. In fact, they may also consult the
opinion of their neighbors before making decisions. For example, it is not
uncommon for people to consider both the recommendation of their peers
(local information) and the stock price (global information) in deciding
which stock to buy from the market. Hence, it is instructive for us to
incorporate the local information into the MG model to gain more insights on
this kind of social and economic systems.

In the past few years, many researchers have studied a few variations of MG
with local information \cite{LMG1,LMG2,LMG3,LMG4}. However, players can only
make use of either the global or local information in each turn in these
models. In contrast, people often make their decisions according to both the
global information and their local information in many social and economic
phenomenon. On the other hand, Quan~\emph{et al.} introduced the local
information in the evolutionary minority game (EMG) \cite{LEMG}. Since we
want to focus on how are players affected by their local information in a
\emph{non-evolutionary} game, we shall not consider the EMG with local
information here. With the above consideration in mind, we would like to
propose a model of MG where players use both the local and global
information in an unbiased manner for making decision.

In Section II, we introduce a new model called the Networked Minority Game
(NMG). It is a modified MG model in which all players can make use of not
only the global information but also the local information from their
neighbors that are disseminated through a network. Results of numerical
simulations are presented and discussed in Section III. Lastly, we conclude
by giving a brief summary of our work in Section IV.

\section{Our Networked Minority Game}

In this Section, we would like to show how to construct the Networked
Minority Game (NMG) model. In this repeated game, there are $N$
heterogeneous inductive reasoning players whose aim is to maximize one's own
profit in the game. In every turn, each player has to choose one out of
$N_c$ alternatives with label $\{0, 1,\ldots,N_c-1\}$. The global minority
choice, denoted as $\Omega_g(t)$ at time $t$, is simply the least popular
choice amongst all players in that turn. (Note that the global minority
choice is the least popular choice that are chosen by a non-zero number of
players; and it is chosen randomly amongst the choices with the least
non-zero number of players in case of a tie.) The players picked the global
minority choice gain one unit of wealth while all the other lose one.

Unlike in MG, not only the global information is delivered to each player as
one's external information in our model. Nevertheless, the global
information given to players are the same in these two models. In NMG, we
call it the global history $(\Omega_g(t-M_g), \Omega_g(t-M_g-1), \ldots,
\Omega_g(t-1))$ which is simply the $N_c$-ary string of the global minority
choice of the last $M_g$ turns. The global history can only take on
$P_g \equiv N_c^{M_g}$ different states. We label these states by an index
$\mu_g = 1, \ldots, P_g$ and denote the global history $(\Omega_g(t-M_g),
\Omega_g(t-M_g-1), \ldots, \Omega_g(t-1))$ by the index $\mu_g(t)$.

\begin{figure}[p]
\includegraphics*[scale = 0.84, bb = 20 510 460 800]{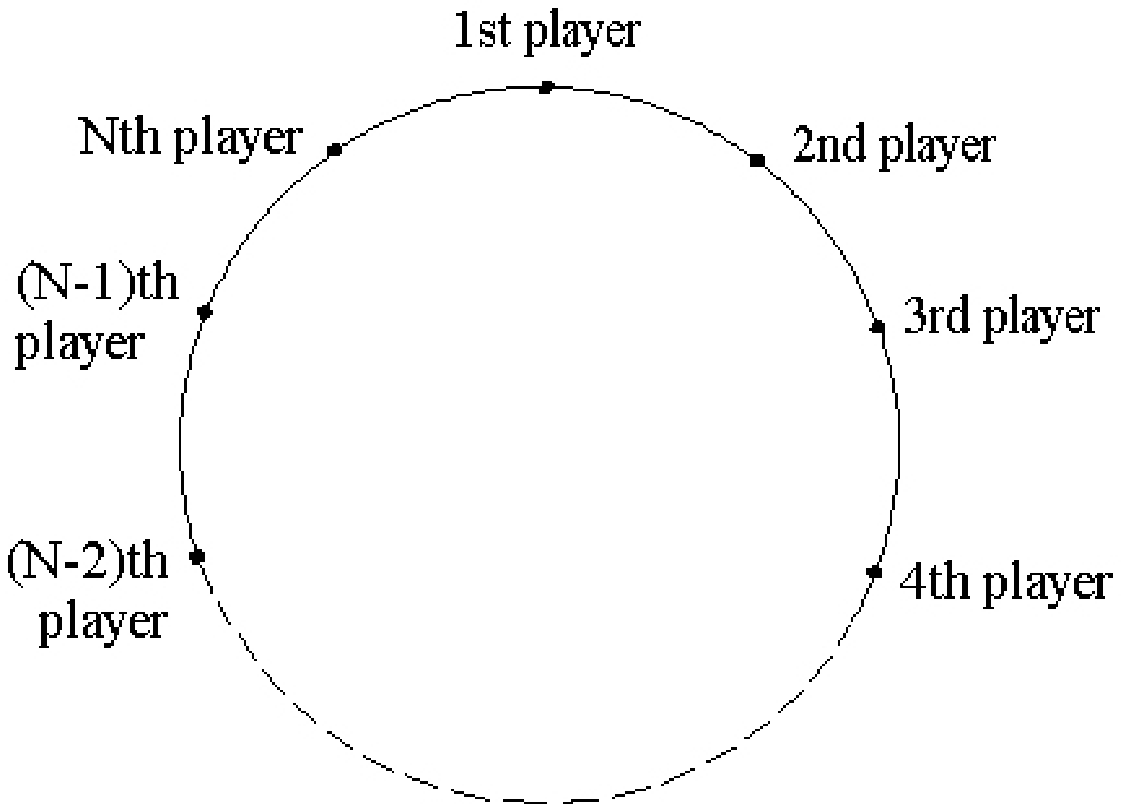}
 \caption{The network of $N$ players on a ring for NMG.}
 \label{fig:f1}
\end{figure}

Besides global information, local information is also distributed via a
network to all players in NMG such that individual player receive one's
local information from a different source. In this game, we arrange all the
players with label $\{ 1, \ldots, N \}$ on a ring (see fig.~\ref{fig:f1})
where the local information of a player is based on the choices of this
player and his/her nearest neighbors on the ring. Specifically, the local
information given to the $i$th player is the so-called local history
$(\Omega_l^i(t-M_l), \Omega_l^i(t-M_l-1), \ldots, \Omega_l^i(t-1))$ which is
simply the $N_c$-ary string of the local minority choice of this player of
the last $M_l \neq 0$ turns. (Note that we do not consider the case of
$M_l = 0$ since it is equivalent to MG.) Here, the local minority choice of
the $i$th player at time step $t$, $\Omega_l^i(t)$, refers to the least
popular choice amongst this player and his/her nearest neighbor on the ring
at this time step. (In other words, $\Omega_l^i(t)$ is the least popular
choice amongst $(i-1)$th, $i$th and $(i+1)$th player.) However, unlike the
global minority choice, the local minority choice could be an alternative
that nobody chooses. In case of a tie, $\Omega_l^i(t)$ is chosen randomly
amongst the choices with the least number of players. The local history can
only take on $P_l \equiv N_c^{M_l}$ different states. We again label these
states by an index $\mu_l = 1,\ldots,P_l$ and denote the local history of
the $i$th player $(\Omega_l^i(t-M_l), \Omega_l^i(t-M_l-1), \ldots,
\Omega_l^i(t-1))$ by the index $\mu_l^i(t)$. By the way, it is easy to
extend NMG to the case where players are connected on a different topology.
For example, we can arrange players on a dynamically random chain in which
each player is connected to two randomly chosen players and all the
connections between players change at every time step.

In brief, each player is given a $N_c$-ary string of length $M = M_l + M_g$
storing both the global and local history to decide his/her choice. Players
can only interact indirectly with each other through the global history
$\mu_g(t)$ and their local history $\mu_l^i(t)$. However, how does each
player make use of such external information to decide one's choice in the
NMG? He/She does so by employing strategies to predict the next minority
choice according to both the global and local information where a strategy
is a map sending individual combination of global and local history
($\mu_g$,$\mu_l$) to the choice $\{0, 1, \ldots, N_c-1\}$. A strategy $s$
can be represented by a vector $\vec{s} \equiv (\chi_s^{(1,1)},
\chi_s^{(1,2)}, \ldots, \chi_s^{(P_g,P_l)})$ where $\chi_s^{(\mu_g,\mu_l)}$
is the minority choice predicted by the strategy $s$ for the input
$(\mu_g,\mu_l)$.

In our model, each player picks $S$ randomly drawn strategies from the
strategy space before the game commences. (We will discuss about the
strategy space in depth later on.) Just like in MG, strategies in NMG are
not evolving, i.~e., players are not allowed to revise their own $S$
strategies during the game. At each time step, each player uses his/her own
best working strategy to guess the next global minority choice. But how does
a player decide which strategy is the best? Players use the virtual score,
which is simply the hypothetical profit for using a single strategy
throughout the game, to evaluate the performance of a strategy. The strategy
with the highest virtual score is considered as the best one.

Since inductive reasoning players do not know whether a strategy is good or
not before the game commences, we cannot restrict players to have ``good
strategies" only. As a result, all strategies in our model must be unbiased
to the input, i.~e.~the global and local history. Thus all the strategies
employed in NMG can be picked from the full strategy space which is
constituted by all the possible strategies for the input. (It is obvious
that these strategies have no bias for any input.) There is a total of
$N_c^{N_c^M}$ distinct strategies in the full strategy space for our model.
Nevertheless, it is very probable that a large number of strategies exist in
this full strategy space even $N_c = 2$. Consequently, we would like to pick
strategies from the reduced strategy space \cite{MG2,MGNu1} to enhance
computational feasibility.

The reduced strategy space is only composed of strategies which are
significantly different from each other. Thus it characterizes the
diversity of strategies in the full strategy space and we can use it instead
of the full strategy space without altering the properties and dynamics of
the game. How can we construct the reduced strategy space for NMG? To answer
this question, let us look at the reduced strategy space of MG first. For
MG, the reduced strategy space consists of mutually uncorrelated and
anti-correlated strategies only. Challet and Zhang showed that the maximal
reduced strategy space of the original 2-choice MG, denoted as $V_2(M)$ for
a 2-choice game with length of global history $M$, is composed of $2^M$
pairs of mutually anti-correlated strategies where any two strategies from
different anti-correlated strategy pairs are uncorrelated with each other
\cite{MG2,MGNu1}. Therefore, there is a total of $2^{M+1}$ different
strategies in $V_2(M)$. (Note that there are other smaller reduced strategy
spaces consisting of less number of strategies for MG \cite{MCMG,RMG,RMG2}.)
It can be shown that, in general, the maximal reduced strategy space
$V_{N_c}(M)$ for $N_c$-choice MG is composed of $N_c^M$ ensembles of
mutually anti-correlated strategies where any two strategies from different
anti-correlated strategy ensembles are uncorrelated with each other
\cite{MCMG,RMG,RMG2}. Moreover, $V_{N_c}(M)$ consists of $N_c^{M+1}$
distinct strategies.

For NMG, it seems reasonable for us to define the reduced strategy space
$R_{N_c}(M)$ to be the maximal reduced strategy space of MG. However, we
find that the cooperative behavior of the players in NMG using $V_{N_c}(M)$
and the full strategy space are greatly different from each other. Indeed,
such a discrepancy is due to the bias of the strategies of $V_{N_c}(M)$ to
some of the input (the global and local history) in our model. Hence, we
should not use $V_{N_c}(M)$ to substitute the full strategy space in NMG.
In fact, we must define the reduced strategy space $R_{N_c}(M)$ in a
different way.

For NMG, a strategy $\vec{s}$ is unbiased to the input if it satisfies the
following conditions:
\newcounter{c}
\begin{list}
  {\arabic{c})}{\usecounter{c}}
\item Simply looking at the predictions of this strategy for a \emph{given}
global history does not give us any information about its predictions for
any other global history.
\item Simply looking at the predictions of this strategy for a \emph{given}
local history does not give us any information about its predictions for any
other local history.
\end{list}
Accordingly, we define the reduced strategy space of NMG as follows:
\begin{eqnarray}
R_{N_c}(M) = \left\{
\begin{array}{ll}
\{ (s_1, s_2, \ldots, s_{P_l}) : s_1, s_2, \ldots, s_{P_l} \in V_{N_c}(M_g)
\} & \mbox{if $M_l \leq M_g$,} \vspace*{1mm} \\
\{ (s_1, s_2, \ldots, s_{P_g}) : s_1, s_2, \ldots, s_{P_g} \in V_{N_c}(M_l)
\} & \mbox{otherwise,}
\end{array} \right. \label{E:RSS}
\end{eqnarray}
where the $i$th player uses the so-called segment strategy $s_j$ to predict
the global minority choice for the global history $\mu_g(t)$ if
$j = \mu_l^i(t)$ in case of $M_l \leq M_g$ and respectively for the local
history $\mu_l^i(t)$ if $j = \mu_g(t)$ in case of $M_l > M_g$. Table
\ref{tab:t1} illustrates how a strategy in $R_{N_c}(M)$ gives the prediction
of the next minority choice. It is easy to show that the number of distinct
strategies in $R_{N_c}(M)$ is either $N_c^{(M_g+1)P_l}$ or
$N_c^{(M_l+1)P_g}$ depending on the ratio of $M_g/M_l$. Note that we should
not define $R_{N_c}(M)$ to be given by just one of the above expressions
over all range of $M_g/M_l$. Otherwise, there is a large redundancy of
strategies in $R_{N_c}(M)$ since many strategies are very similar for such
case. (However, it does not matter if we define $R_{N_c}(M)$ by the second
expression when $M_l = M_g$.) Indeed, we have verified that the cooperative
behavior of the players in NMG using $R_{N_c}(M)$ and the full strategy
space agree well with each other; i.~e., $R_{N_c}(M)$ can successfully
characterize the diversity of strategy in the full strategy space.

\begin{table}[ht]
 \begin{tabular}{@{\extracolsep{5mm}}ccc}
  \hline \hline
  \rule{0pt}{0.15in} global history $\mu_g$ & local history $\mu_l$ &
prediction $\chi_s^{(\mu_g,\mu_l)}$ \\ \hline
  \rule{0pt}{0.15in} (0,0) & (0) & 1 \\
  \rule{0pt}{0.15in} (0,1) & (0) & 1 \\
  \rule{0pt}{0.15in} (1,0) & (0) & 1 \\
  \rule{0pt}{0.15in} (1,1) & (0) & 1 \\
  \rule{0pt}{0.15in} (0,0) & (1) & 0 \\
  \rule{0pt}{0.15in} (0,1) & (1) & 0 \\
  \rule{0pt}{0.15in} (1,0) & (1) & 1 \\
  \rule{0pt}{0.15in} (1,1) & (1) & 1 \\
  \hline \hline
 \end{tabular}
 \caption{The prediction of the minority choice $\chi_s^{(\mu_g,\mu_l)}$ by
the strategy $s = (s_1, s_2)$ where $\vec{s}_1 \equiv (1,1,1,1)$ and
$\vec{s}_2 \equiv (0,0,1,1)$ for $M_g = 2$ and $M_l = 1$.}
 \label{tab:t1}
\end{table}

To investigate how well players cooperate with each other in our game, a
quantity of interest is the attendance of an alternative $A_j(t)$ which is
the number of players choosing the alternative $j$ at time $t$. For players
gaining the maximum profit, the expectation value of the attendance of any
alternative should be equal to $N/N_c$. Accordingly, the variance of the
attendance $\sigma_j^2 = \langle (A_j(t))^2 \rangle - \langle A_j(t)
\rangle^2$ represents the loss of players in the game. (Here, $\langle \;
\rangle$ denotes the average over time.) Hence, we would like to study
$\langle A_j(t) \rangle$ and $\sigma_j^2$ as a function of the complexity of
the system for our model. Which parameter can be used as a measure of the
complexity of the system? It is the so-called control parameter $\alpha$
\cite{MG2,MGNu1,MGNu2} which is the ratio of the strategy space size to the
number of strategies at play. For NMG, the control parameter $\alpha$ is
equal to either $N_c^{(M_g+1)P_l}/NS$ for $M_l \leq M_g$ or
$N_c^{(M_l+1)P_g}/NS$ for $M_l > M_g$.

On the other hand, we also want to compare the variance of the attendance
with the prediction by the crowd-anticrowd theory \cite{CAC1,CAC2,CAC3}
in order to investigate the crowding effect in our model. Since the
strategies used in NMG (which are picked from $R_{N_c}(M)$) are neither
anti-correlated nor uncorrelated, we cannot only consider the
interactions of the anti-correlated strategies in the crowd-anticrowd
calculation of the variance just like in MG \cite{CAC2,CAC3} and the
multichoice MG \cite{RMG}. In fact, each strategy used in NMG is a set of
segment strategies from either $V_{N_c}(M_l)$ or $V_{N_c}(M_g)$ whereas
different segment strategies are used for different $\mu_l^i(t)$ or
$\mu_g(t)$ (see Eq.~(\ref{E:RSS})). Thus we can apply the crowd-anticrowd
theory to estimate the variance simply by counting the crowd-anticrowd
cancellation of the anti-correlated segment strategies. Accordingly, the
crowd-anticrowd prediction of the variance in NMG is given as follows:
\begin{equation}
\sigma^2 = \left\{
\begin{array}{ll}
\left\langle \displaystyle \frac{1}{N_c} \sum_{\mathcal{S}_g \in
V_{N_c}(M_g)} \sum_{a \in \mathcal{S}_g} \left\{ \frac{1}{N_c^2} \left[
\sum_{b \in \mathcal{S}_g \backslash \{a\}} (N_{a} - N_{b}) \right]^2
\right\} \right\rangle & \mbox{if $M_l \leq M_g$,} \vspace*{1mm} \\
\left\langle \displaystyle \frac{1}{N_c} \sum_{\mathcal{S}_l \in
V_{N_c}(M_l)} \sum_{a \in \mathcal{S}_l} \left\{ \frac{1}{N_c^2} \left[
\sum_{b \in \mathcal{S}_l \backslash \{a\}} (N_{a} - N_{b}) \right]^2
\right\} \right\rangle & \mbox{otherwise,}
\end{array} \right.
\label{E:CAC_var}
\end{equation}
where $\mathcal{S}_g$ and $\mathcal{S}_l$ denotes the mutually
anti-correlated strategy ensembles in $V_{N_c}(M_g)$ and $V_{N_c}(M_l)$
respectively, and $N_{a}$ is the number of players making decision according
to the segment strategy $a$. We should aware that the variance of attendance
for different alternatives must equal when averaged over time and initial
choice of strategies since there is no bias for any alternative in our game.

\section{Results of the game}

In all the simulations reported in this paper, each set of data was recorded
from $1000$ independent runs. In each run, we run 10000 steps starting from
initialization before making any measurements. We have checked that it is
already enough for the system to attain equilibrium in all the cases
reported here. Then we took the average values on 15000 steps after the
equilibration. To be computational feasible, we choose the case where
$N_c = 2$, $S = 2$ and $M = M_l + M_g = 5$.
  
\subsection{Comparison of NMG with MG}

In this section, we compare the performance of players in NMG with that in
MG. Since the system exhibits peculiar properties in NMG with $M_g = 0$, we
delay the discussion about this case to Section~\ref{Mg0}.

Let us begin by investigating the properties of the mean attendance as a
function of the control parameter $\alpha$ in NMG. Obviously, the mean
attendance in NMG is similar to MG for all values of $M_l/M_g$. That is
to say, it always fluctuates around $\lfloor N/2 \rfloor$ such that players
can maximize their global profit. 

\begin{figure}[ht]
\includegraphics*[scale = 0.7, bb = 25 0 726 812]{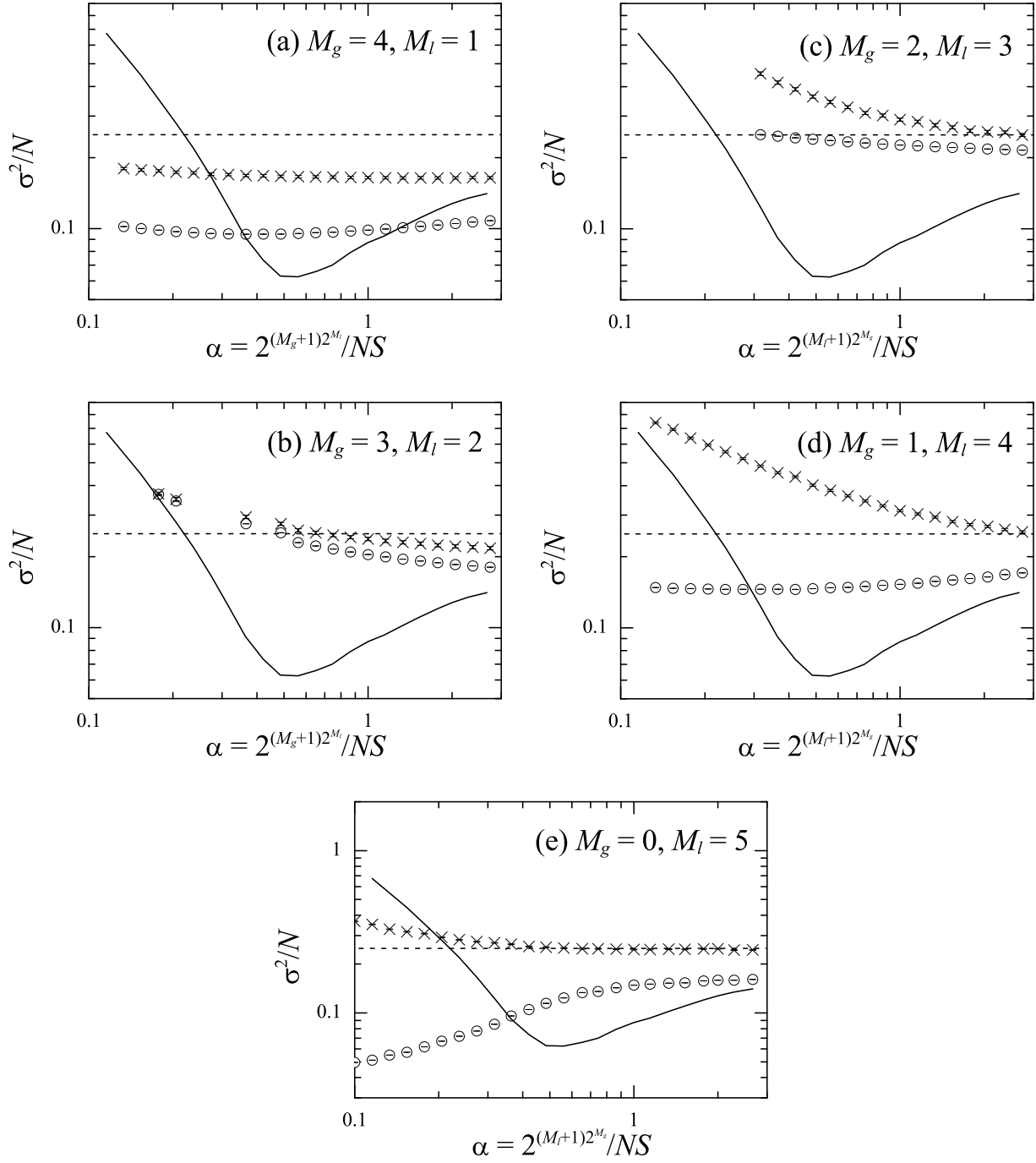}
 \caption{The variance of the attendance per player $\sigma^2/N$ (opaque
circles) versus the control parameter $\alpha$ for NMG with $N_c = 2$,
$S = 2$ and $M = M_g + M_l = 5$ where players are connected on a ring. The
crosses are the predictions of the crowd-anticrowd theory whereas the dashed
lines indicate the coin-toss value, i.~e., the corresponding value in the
random choice game. For comparison purpose, the solid lines indicate the
corresponding numerical results in the original MG.}
 \label{fig:f2}
\end{figure}

Next we evaluate the performance of the players in NMG by studying the
variance of the attendance per player $\sigma^2/N$ versus the control
parameter $\alpha$ as shown in fig.~\ref{fig:f2}. (Note that the variance of
the two choices must be the same by symmetry for a 2-choice game; and all
the variance mentioned in this section are divided by the number of player
$N$ for objective comparison.) We find that the variance in NMG is always
much smaller than that in MG for small $\alpha$ no matter what is the value
of $M_g/M_l$. In other words, players perform much better in NMG than in MG
through the introduction of their own local information when $\alpha$ is
small. When the reduced strategy space size is relatively small comparing
with the number of strategies at play, the fluctuation of the attendance is
large in MG due to the overcrowding effect of player's strategies
\cite{MGNu2}. Under the overcrowding effect, players tend to choose the same
alternative for the same global history since they use some very similar
strategies. However, in NMG, players can choose different alternatives even
they use the same strategy since they consider both their local information
and the global information in deciding their choice. Such phenomenon will
dilute the overcrowding effects of the strategies in NMG and so players can
perform much better than in MG when $\alpha$ is small. 

When the control parameter $\alpha$ increases, the overcrowding effect will
be suppressed in both NMG and MG. In MG, the maximal cooperation amongst the
players can be achieved subsequently when the number of strategies at play
is approximately equal to the reduced strategy space size. However, in NMG,
players need to cooperate with each other through both the global
information and their local information. These mixed types of information
obtained from different sources makes the cooperation amongst all players
becomes much more difficult for all value of $M_g/M_l$. So the variance in
NMG is larger than in MG around the critical point $\alpha_c$ and it tends
to the coin toss value (which is the variance resulting fromplayers making
random choices throughout the game) when $\alpha$ increases further. 

\subsection{NMG with global information (i.~e.~$M_g \neq 0$)}

Fig.~\ref{fig:f2} shows the variance of attendance per player $\sigma^2/N$
as a function of the control parameter $\alpha$ in NMG for different ratio
of $M_g$ to $M_l$. When $M_l < M_g$, the variance becomes larger as $M_l$
increases regardless of the value of the control parameter $\alpha$.
Moreover, the variance tends to the coin-toss value for $M_l \approx M_g$
over all range of $\alpha$. Furthermore, when $M_l > M_g$, the performance
of players becomes better if more local information is available no matter
what is the value of $\alpha$. To account for this phenomenon, we should
consider the structure of strategies in the reduced strategy space
$R_{N_c}(M)$. When $M_l < M_g $, each strategy is composed of more segment
strategies belonging to a smaller strategy space $V_{N_c}(M_g)$ as $M_l$
increases. Therefore, players are more likely to use similar strategies and
the overcrowding effect of strategies will dominate which results in large
fluctuations of the attendance. Similarly, when $M_l > M_g$, the segment
strategies are drawn from $V_{N_c}(M_l)$ and thus the variance becomes
larger as $M_g$ increase due to the overcrowding effect of strategies. In
particular, when $M_l \approx M_g$, each strategy is composed of the largest
number of segment strategies drawn from the smallest strategy space
($V_{N_c}(M_l)$ or $V_{N_c}(M_g)$) and so the variance is the largest for
such case.

\subsection{NMG with no global information (i.~e.~$M_g = 0$)} \label{Mg0}

\begin{figure}[ht]
\includegraphics*[scale = 0.45, bb = 45 0 730 580]{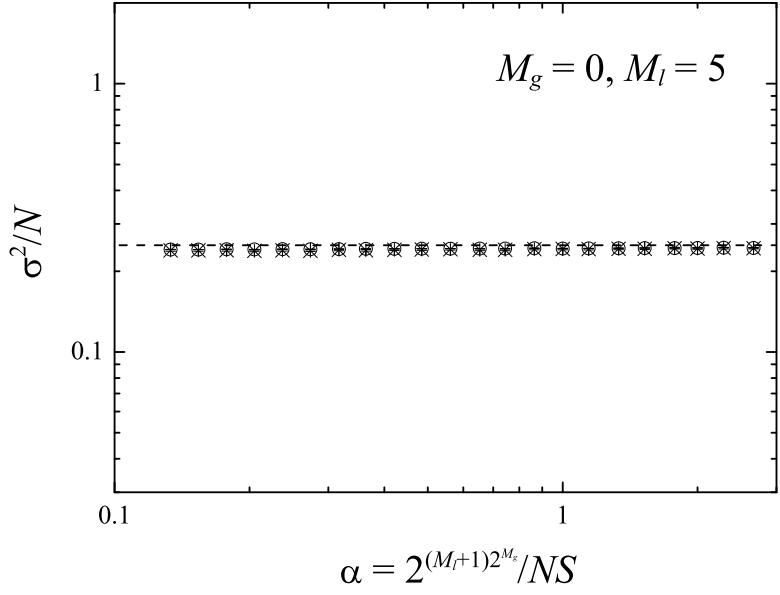}
 \caption{The variance of the attendance per player $\sigma^2/N$ (opaque
circles) versus the control parameter $\alpha$ for NMG with $N_c = 2$,
$S = 2$ and $M = M_g + M_l = 5$ where players are connected on a dynamically
random chain. The crosses are the predictions of the crowd-anticrowd theory
whereas the dashed lines indicate the coin-toss value.}
 \label{fig:f3}
\end{figure}

When $M_g = 0$, only local information is available to each player. From
numerical simulation, the variance in NMG with $M_g = 0$ is found to be
smaller than in MG and NMG with $M_g \neq 0$ when the control parameter
$\alpha$ is small and approaches zero. It implies that the cooperation
amongst players in NMG with $M_g = 0$ is much better than in MG and NMG with
$M_g \neq 0$ for small $\alpha$. In fact, player's cooperation in NMG with
no global information is resulted from their local interaction. To
investigate the local interaction, we calculate the correlation function for
the local minority choices of two neighboring players as follows:
\begin{equation}
 \rho_{i-1,i} = \frac{\langle \Omega_l^{i-1}(t) \Omega_l^{i}(t) \rangle -
\langle \Omega_l^{i-1}(t) \rangle \langle \Omega_l^{i}(t) \rangle}{\langle
[ \Omega_l^{i}(t) ]^2 \rangle - \langle \Omega_l^{i}(t) \rangle^2},
\end{equation}
where $\Omega_l^{i}(t)$ is the local minority choice of the $i$th player at
time $t$. We reveal that the correlation function $\rho_{i-1,i}$ is equal to
infinity for many neighboring players when $M_g = 0$. That is to say, many
players always choose the same alternative as their nearest neighbors
throughout the game. In such case, we found that the local histories of
these players remain the same such that their local minority choices are
completely frozen. In fact, the decision of a frozen player is dominated by
the components of segment strategies corresponding to $\mu_g = (0,\ldots,0)$
or $\mu_g = (1, \ldots, 1)$ whenever $M_g = 0$. For instance, suppose the
components of segment strategies corresponding to $\mu_g = (0,\ldots, 0)$
equal $1$ for all $S$ strategies of two nearest neighbors. Once their local
histories are $(0, \ldots, 0)$, their own local minority will then be $0$ in
this turn and their local histories will be $(0, \ldots, 0)$ again at the
next time step. It is a fixed point of the dynamics leading to the dimension
reduction of the reduced strategy space. If similar configurations arise
along the ring, the choices of some players can be determined before each
turn because their actions are frozen. The freezing effect of players will
minimize the fluctuations of attendance since it is only contributed by few
non-frozen players. Indeed, it is the negligence of global information that
allows the strong local interactions amongst players who are connected on a
ring. On the contrary, players can only weakly interact with each other
locally if they are placed on a dynamically random chain in which each
player is connected to two randomly chosen players with all the connections
between players change at every time step. Hence, whenever $M_g = 0$, the
freezing effect of players disappears and the variance tends to the
coin-toss value for the case of the dynamically random chain as shown in
fig.~\ref{fig:f3}.

On the other hand, numerical results show that the probability for the
freezing of players decreases exponentially when $M_g$ increases for
$M_l > M_g$. It is because each strategy is composed of more number of
segment strategies and the disturbance from the global information becomes
more significant when more global information is available. So the variance
becomes larger as $M_l$ decreases for $M_l > M_g$. 

\subsection{Comparison of the numerical results of NMG with the predictions
by the crowd-anticrowd theory}

\begin{figure}[ht]
\includegraphics*[scale = 0.45, bb = 45 0 730 580]{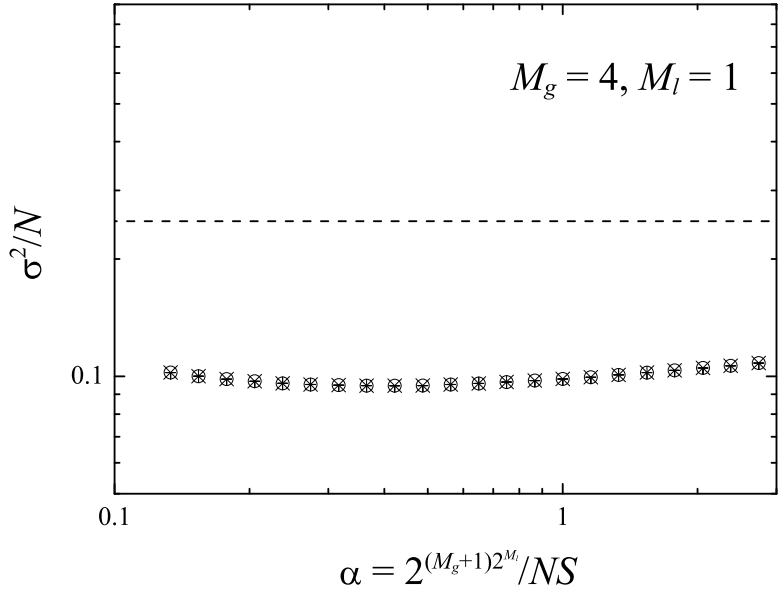}
 \caption{The variance of the attendance per player $\sigma^2/N$ (opaque
circles) versus the control parameter $\alpha$ for NMG with $N_c = 2$,
$S = 2$ and $M = M_g + M_l = 5$ where players are connected on a ring. The
crosses are the corrected semi-analytical results predicted by the
crowd-anticrowd theory (by counting the contribution of both the
anti-correlated and uncorrelated segment strategies to the crowd-anticrowd
cancellation) whereas the dashed lines indicate the coin-toss value.}
 \label{fig:f4}
\end{figure}

In order to investigate the crowding effect in NMG, we compare the numerical
results of the variance with the predictions by the crowd-anticrowd theory.
We find a large discrepancy between them whenever $M_g \gg M_l$ or
$M_l \gg M_g$ although their trends are consistent with each other for all
value of $M_l/M_g$. According to the crowd-anticrowd theory, the variance of
attendance for a given realization of the quenched disorder $\Omega$
(i.~e.~the initial configuration of the system) is given by \cite{CAC3}
\begin{equation}
 \sigma^2_\Omega = \sum_{R} \langle (a_R^{\mu(t)})^2  (n_R(t))^2 \rangle +
\sum_{R} \langle a_R^{\mu(t)} a_{\bar{R}}^{\mu(t)} n_R(t) n_{\bar{R}}(t)
\rangle + \sum_{R \neq R^{'} \neq \bar{R}} \langle a_R^{\mu(t)}
a_{R^{'}}^{\mu(t)} n_R(t) n_{R^{'}}(t) \rangle,
\end{equation}
where $\sum_R$ denotes the sum over the strategies in the corresponding
reduced strategy space, $a_R^{\mu(t)}$ is the action of strategy $R$ to the
history $\mu$ at time $t$ ($a_R^{\mu(t)} = \pm 1$ corresponds to the two
alternatives) and $n_R(t)$ is the number of players using strategy $R$ at
time $t$. In MG, $\sum_{R \neq R^{'} \neq \bar{R}} \langle a_R^{\mu(t)}
a_{R^{'}}^{\mu(t)} n_R(t) n_{R^{'}}(t) \rangle$ should be equal to zero and
thus is dropped. (For MG, the variance in fact plays the role of the
effective Hamiltonian of the spin-glass-like system while the strategies can
be interpreted as the quenched disorder \cite{MGHam1,MGHam2}; thus we can
state that the four-point correlation function is dropped in the effective
Hamiltonian of the system for MG. ) Such expectation is based on the fact
that the number of players using an uncorrelated strategy pair are, on
average of the history $\mu$, independent from the response of the strategy
pair to the history in MG. However, from numerical simulation, we find that
$\sum_{R \neq R^{'} \neq \bar{R}} \langle a_R^{\mu(t)} a_{R^{'}}^{\mu(t)}
n_R(t) n_{R^{'}}(t) \rangle$ is equal to a large negative number for NMG
with $M_g \gg M_l$. In other words, the uncorrelated segment strategy pairs
are no longer independent with each other. Through the local interaction,
players tend to use the uncorrelated segment strategy pairs whenever the
pairs choose the opposite choice. It is a favorable solution of the dynamics
because the uncorrelated segment strategies can now contribute to the
crowd-anticrowd cancellation which will in turn lead to the further
reduction of the fluctuation of attendance. In fact, the crowd-anticrowd
cancellation by the uncorrelated strategies is only possible in NMG since
players have the freedom to choose which uncorrelated segment strategies to
be used on the basis of the local information. We can correct the
semi-analytical results of the crowd-anticrowd theory by counting the
contribution of both the anti-correlated and uncorrelated segment strategies
to the crowd-anticrowd cancellation. The corrected semi-analytical results
is found to match with the numerical one as shown in fig.~\ref{fig:f4}. So
we conclude that the discrepancy is mainly due to the crowd-anticrowd
cancellation by the uncorrelated strategies for $M_g \gg M_l$. 

When $M_l \gg M_g$, the discrepancy of the predictions of the
crowd-anticrowd theory from the numerical results is due to the freezing of
players. The ergodicity of the local histories of frozen players is broken
down under the freezing effects. In other words, some possible states of the
local histories will never be visited for frozen players since the effective
dimension of the strategy space is reduced. Of course, this will violate the
ergodicity assumption in the crowd-anticrowd theory. So the results
predicted by the crowd-anticrowd theory show a large discrepancy from the
numerical results when $M_l \gg M_g$. 

\section{Conclusion}

In summary, we find that NMG exhibits a remarkably different behavior from
MG. In NMG with non-zero global information, selfish players on the ring
can cooperate with each other to reduce their loss; moreover, their
cooperation is much better than in MG by using the local information that
are disseminated through the ring except when the number of strategies at
play is approximately equal to the strategy space size. Such phenomenon is
believed to be due to the dilution of the crowding effect of players which
is resulted from their local interactions on the ring. 

In NMG with no global information, many players on the ring are found to
be frozen where their action and also their local minority choice remain the
same throughout the game. We reveal that the freezing occurs when their
decisions are dominated by the components of segment strategies
corresponding to $\mu_g = (0, \ldots, 0)$ or $\mu_g = (1, \ldots, 1)$.
Such domination arises because of the strong local interaction on the
ring due to the absence of the global information.

On the other hand, we find that the predictions of the crowd-anticrowd
theory deviate very much from the numerical results for NMG. Such
discrepancy is found to be due to the crowd-anticrowd cancellation
contributed by the uncorrelated strategies in NMG whereas it is impossible
for the uncorrelated strategies to contribute to the crowd-anticrowd
cancellation in MG.

In fact, all the above arguments should still be valid when players are
connected on a network with different topology in NMG provided that each
player always obtains the local information from the same neighbors; and
thus we believe that the cooperative behavior of players are similar in
all these cases.

Finally, we would like to point out that the order parameter is also worth
to be studied for NMG. From this study, we may learn whether there is a
phase transition from a symmetric phase to an asymmetric phase as the
complexity of the system increases just like that in MG.

\end{document}